\documentclass[aps,pre,two column,superscriptaddress]{revtex4}
\usepackage{amssymb,amsfonts,amsmath}
\usepackage{graphics}
\usepackage{graphicx}
\usepackage{setspace}
\usepackage{subfigure}

\begin{document}

%\preprint{}

\title{Equation of state of metallic hydrogen from Coupled Electron-Ion
Monte Carlo simulations}


\author{Miguel A. Morales}
\email[]{mmorale3@illinois.edu}
%\thanks{}
\affiliation{Department of Physics, University of Illinois Urbana-Champaign}

\author{Carlo Pierleoni}
\email[]{carlo.pierleoni@aquila.infn.it}
%\thanks{}
\affiliation{Consorzio Nazionale Interuniversitario per le
Scienze Fisiche della Materia and Physics Department,
University of L'Aquila, Italy}

\author{David Ceperley}
\email[]{ceperley@illinois.edu}
%\thanks{}
\affiliation{Department of Physics, National Center of
Supercomputer Applications, and Institute of Condensed Matter
Physics, University of Illinois Urbana-Champaign}


\date{\today}

\begin{abstract}
We present a study of hydrogen at pressures higher than
molecular dissociation using the Coupled Electron-Ion Monte
Carlo method. These calculations use the 
accurate Reptation Quantum Monte Carlo method to estimate the
electronic energy and pressure while doing a Monte Carlo
simulation of the protons. In addition to presenting simulation
results for the equation of state over a large region of phase
space, we report the free energy obtained by thermodynamic
integration. We find very good agreement with DFT calculations
for pressures beyond 600 GPa and densities above $\rho=1.4
g/cm^3$. Both thermodynamic as well as structural properties
are accurately reproduced by DFT calculations. This agreement
gives a strong support to the different approximations employed
in DFT, specifically the approximate exchange-correlation
potential and the use of pseudopotentials for the range of
densities considered. We find disagreement with chemical
models, which suggests a reinvestigation of planetary models,
previously constructed using the Saumon-Chabrier-Van Horn
equations of state.
\end{abstract}

%\keywords{}

\maketitle

\section{Introduction}
%\subsection{}
%\subsubsection{}

Although hydrogen is the first element in the periodic table,
its phase diagram has diverse phases. As the most abundant
element in the universe, it is important to have an accurate
understanding of its properties for a large range of pressure
and temperature. A qualitative description is not sufficient
because; for example, models of hydrogenic planets require
accurate results to make correct predictions \cite{Guillot05}.

The high pressure phases of hydrogen have received considerable
attention in recent years,  both from theory and experiment. At
lower temperatures, static compression experiments using
diamond anvil cells can reach pressures of 320 GPa, where the
quest to find the metal-insulator  and molecular-atomic
transitions in the solid phase still continues
\cite{Loubeyre02}. Dynamic compression experiments using either
isentropic compression or shock waves, are used at higher
temperatures and can now reach pressures above 200 GPa
\cite{Hicks09,Nellis06}. Even though experimental techniques at
high pressure have improved considerably over the last decade,
they are still not accurate enough to provide conclusive
answers to many of the relevant questions. Although this
situation might change in the near future with the construction
of more powerful machines such as the National Ignition
Facility, computer simulations today provide the most reliable
method for determining the thermodynamic properties at high
pressures and temperatures.

Many theoretical techniques have been used including: free
energy minimization methods in the chemical picture
\cite{Saumon95,Kerley03,Redmer06}, restricted Path Integral
Monte Carlo (PIMC) \cite{Militzer02} and density functional
theory (DFT) based molecular dynamics (MD)
\cite{Lenosky00,Dejarlais03,Scandolo03,Bonev04_1,Bovev04_2,Militzer07,Redmer08}.
All of these methods employ different approximations that can
affect properties in ways that are difficult to quantify due to
the lack of conclusive experimental results. While free energy
methods are typically accurate in the molecular phase at low
pressures, where molecules are tightly bound and there are
enough experimental results to produce accurate empirical
potentials, at higher density, with the onset of dissociation
and metallization in the liquid, they become unreliable.
Restricted PIMC, on the other hand, is accurate at very high
temperatures where the nodes of the density matrix are known,
but at temperatures below approximately 20,000K its accuracy
(and efficiency) has been limited. For intermediate
temperatures and high pressures, DFT has become the
computational method of choice over the last decade, mainly due
to its advanced development stage and easy accessibility with
many available codes. Practical implementations of DFT employ
pseudopotentials and approximate exchange-correlation
functionals and do not typically estimate quantum proton
effects; these approximations limit its accuracy and
applicability, especially at high pressures. Despite its
possible limitations, DFT is state of the art in \emph{ab
initio} simulations. For example, planetary models are being
built with its equation of state, superseding the well known
Saumon-Chabrier-Van Horn (SCVH) multiphase equation of state
\cite{Saumon95,Militzer08}.

Because of its wide spread use and potential impact in the near
future, it is important to test the validity of the DFT
approximations at extreme conditions, and to determine its
range of applicability. In order to obtain more accurate
results, especially at intermediate temperatures, we need to
employ methods that can go beyond the usual single-body mean
field approximations typically used. Quantum Monte Carlo (QMC)
is a perfect candidate for the task, presenting a good balanced
between speed and accuracy. It does not rely on
pseudopotentials, and correlation effects are treated
explicitly in the full many-body problem. At present, QMC has
the potential to be more accurate for electronic properties
than DFT while being considerably less expensive than quantum
chemistry methods \cite{Umrigar07,Rios06,Grossman00}.

Coupled Electron-Ion Monte Carlo (CEIMC) is a QMC based
\emph{ab initio} method developed to use QMC electronic energy
in a Monte Carlo simulation of the ionic degrees of freedom
\cite{Dewing02,Pierleoni06}. Thanks to recent advances in QMC
methodology we can now obtain results with small systematic
errors. Specifically, the use of Twist Averaged Boundary
Conditions (TABC) \cite{Lin01} together with recently developed
finite-size correction schemes \cite{Chiesa06,Drummond08} allow
us to produce energies that are well converged to the
thermodynamic limit with $\sim$100 atoms; see appendix
\ref{app:sizeef} for additional details. Improvements in the
wavefunctions used for high pressure hydrogen allow us to get
very accurate results \cite{Pierleoni08} and avoid the main
limitations of previous applications of the CEIMC method.

The impressive increase in computational power in recent years
made possible a comprehensive study of the equation of state of
hydrogen. In this paper we present the results of free energy
calculations of hydrogen in its atomic liquid phase for
pressures beyond 150 GPa.  In section \ref{sec:CEIMC}, we
present a brief description of the CEIMC method. In section
\ref{sec:EOS} and appendix \ref{app:eos1} we present the
equation of state (EOS), including the free energy
calculations. In section \ref{sec:Compare} we present a
comparison with other methods, with a special attention to
DFT-based MD results, while in section \ref{sec:Conclusion} we
draw some conclusions. Finally in appendices \ref{app:sizeef}
and \ref{app:convergence} we describe details of the method to
estimate the finite size corrections and of the RQMC
implementation, respectively.

\section{Coupled Electron-Ion Monte Carlo Method}
\label{sec:CEIMC}

CEIMC, in common with the large majority of \emph{ab initio}
methods, is based on the Born-Oppenheimer separation of
electronic and ionic degrees of freedom. In addition, the
electrons are considered to be in their ground state, for some
particular arrangement of the protons. Protons, either
considered as classical or quantum particles
\cite{Pierleoni04}, are instead assumed to be at thermal
equilibrium with a heat bath at a temperature $T$. In the
present calculation, the system of $N$ protons and $N$
electrons is enclosed in a fixed volume $V$ at a number density
$n=N/V$, which we often express with the parameter $r_s=(3/4\pi
n)^{(1/3)}$. The mass density is related to $r_s$ by: $\rho=(3
m_h)/(4 \pi r_s^3)$, with $m_h$ the mass of a hydrogen atom.

We start from the non-relativistic Hamiltonian
\begin{equation}
\hat{H}=-\sum_{i=1}^{2N}\lambda_{i}\nabla_{i}^{2} +
\frac{e^{2}}{2}\sum_{i\neq j}
\frac{z_{i}z_{j}}{|\hat{\vec{r}}_{i}-\hat{\vec{r}}_{j}|}
\label{eq:hamiltonian}
\end{equation}
 $z_{i}$, $m_{i}$, $\hat{\vec{r}}_{i}$ represent respectively
the valence, mass and position operators of particle $i$, and
$\lambda_{i}=\hbar^{2}/2m_{i}$. Let us denote with
$R=(\vec{r}_{1},\cdots,\vec{r}_{N})$ and
$S=(\vec{r}_{N+1},\cdots,\vec{r}_{2N})$ the set of coordinates
of all electrons and protons respectively. \footnote{We
restrict the discussion to spin-unpolarized systems, i.e.,
systems with a vanishing projection of the total spin along any
given direction, say $S_{z}=0$. }

Within the Born-Oppenheimer approximation, the energy of the
system for a given nuclear state $S$ is the expectation value
of the hamiltonian $\hat{H}$ over the corresponding exact ground
state $\left|\Phi_{0}(S)\right>$
\begin{equation}
E_{BO}(S)={\left<\Phi_{0}(S)\left|\hat{H}\right|\Phi_{0}(S)\right>},
\end{equation}
which is a $3N$-dimensional integral over the electron
coordinates in configuration space
\begin{eqnarray}
E_{BO}(S) & = & \int dR~\Phi_{0}^{*}(R|S) \hat{H}(R,S) \Phi_{0}(R|S)  \nonumber \\
 & = & \int dR \left|\Phi_{0}(R|S)\right|^{2} E_{L}(R|S),
\label{eq:Ebo}
\end{eqnarray}
with the \textit{local energy} defined as
\begin{equation}
E_{L}(R|S)=\frac{\hat{H}(R,S)\Phi_{0}(R|S)}{\Phi_{0}(R|S)}.
\label{eq:Eloc}
\end{equation}
In this work, we use Reptation Quantum Monte Carlo (RQMC)
\cite{Baroni99} with the bounce algorithm \cite{Pierleoni05} to
solve the electronic problem.

With the ability to compute the Born-Oppenheimer electronic energy, the Metropolis
algorithm is able to generate a sequence of ionic states according to the Boltzmann
distribution $P(S) \propto e^{-\beta E_{BO}(S)}$ at the inverse temperature $\beta$.
In CEIMC the estimate of $E_{BO}(S)$ for a given trial function
is computed by QMC and it is therefore affected by statistical
noise which, if ignored, will  bias the result. In the Penalty Method\cite{Ceperley99} we require detailed
balance to hold on average (over the noise distribution).
The noise on the energy difference
causes extra rejection with respect to the noiseless case.

The accuracy of the calculations depend crucially on the choice
of wavefunction. The Slater-Jastrow  wavefunction has the form:
 \begin{equation}
\Psi_T(R,S)=D_{\uparrow}D_{\downarrow}e^{-U},
\end{equation}
where U is the sum over all distinct pairs of particles of the
RPA-Jastrow function \cite{Kwon93}. The orbitals in the Slater
determinant are obtained from a DFT calculation in a planewave
basis, using the local density approximation for the
exchange-correlation functional, as parameterized by Perdew,
et. al. \cite{Perdew81,Ceperley80}. The resulting orbitals are
transformed to a cubic spline basis from which they are
interpolated in the QMC calculations. The use of a spline basis
in the RQMC runs represents a large increase in the efficiency
of the simulations. To reduce the computational overhead
produced by the DFT calculations, which must be converged to
self-consistency for every ionic step, we perform a single
self-consistent DFT calculation at the gamma point. Using the
resulting electron density, we build and diagonalize the
Kohn-Sham hamiltonian at the specific points in the Brillouin
zone used in the TABC calculations.

We apply a backflow transformation to the electron coordinates
in the Slater determinant; this introduces correlations and
improves the nodal surfaces of the DFT orbitals. The form of
the transformation is:
\begin{equation}
  \vec{x}_i = \vec{r}_i + \sum_j \eta_{ij}\left(\left|r_{ij}\right|\right)\vec{r}_{ij},
\end{equation}
where $\eta$ is either a parameterized  electron-electron
backflow function or an analytic forms derived using Bohm-Pines
collective coordinates approach \cite{Holzmann03}. As shown in
a recent publications \cite{Pierleoni08,Rios06}, the use of
backflow transformations improves the trial wavefunction,
especially for small projection times. The use of DFT orbitals
with RPA-derived Jastrow and backflow functions represent a
good balance between accuracy and efficiency.

\section{Equation of State of Hydrogen}
\label{sec:EOS}

\begin{figure}[tbp]
 \includegraphics[scale=0.65]{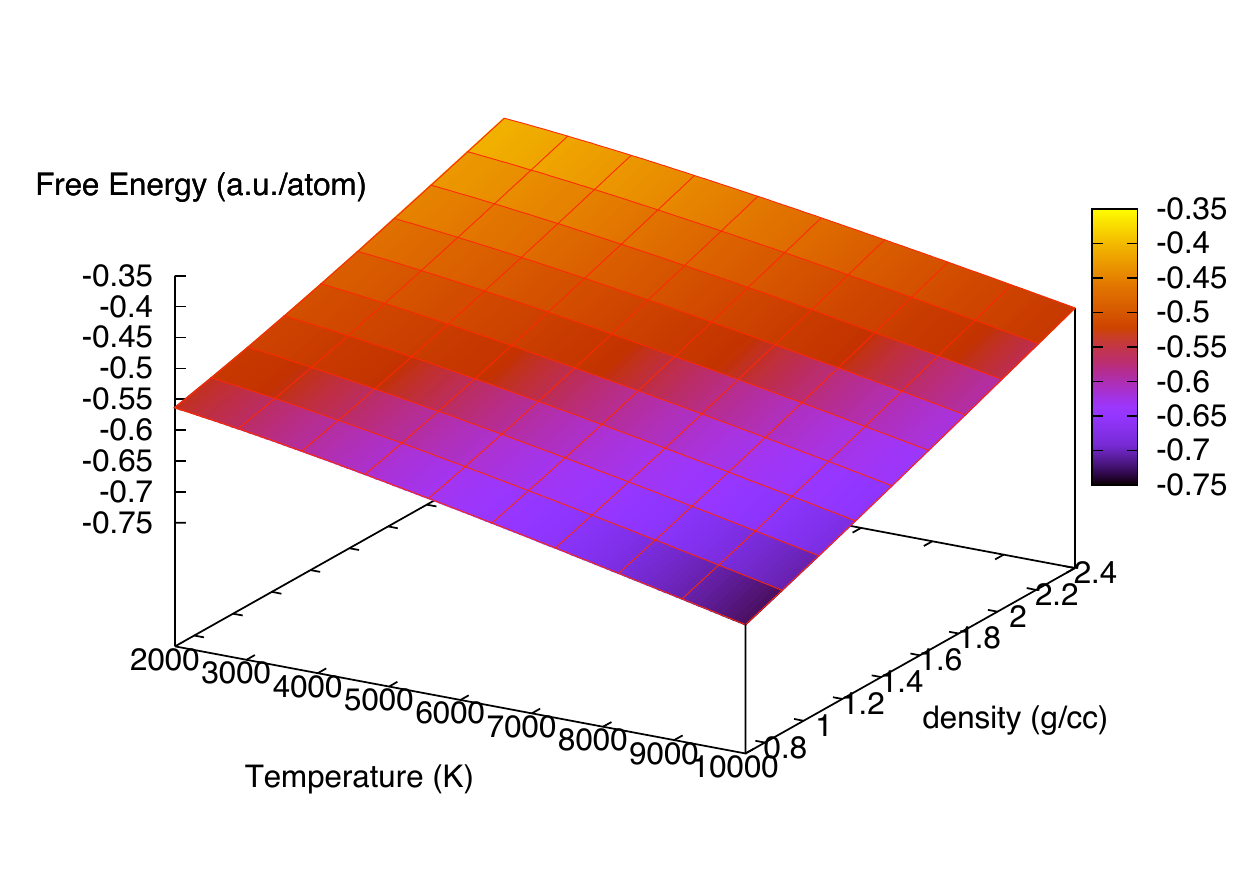}
 \caption{\label{fig1} Free energy surface of hydrogen as a function of temperature and density. }
\end{figure}

In this paper we calculate the free energy of hydrogen as a
function of temperature and density. We explore the range 2
000K $<$ T $<$ 10 000K and 0.7 g cm$^{-3}$ $<  \rho <$ 2.4 g
cm$^{-3}$. For $\rho$ $<$ 0.7 g cm$^{-3}$, molecular
dissociation becomes the dominant feature in the EOS, requiring
a more detailed study than the one performed here  to reach a
similar level of accuracy. Simulations of the molecular liquid
are in progress.

The CEIMC calculations were performed with 54 hydrogen atoms
using RQMC. The time step and projection length used in RQMC
were chosen to reach a convergence of the energy of 0.2-0.3
mHa/atom; see appendix \ref{app:convergence} for details. We
used TABC with a grid of 64 twists (96 for $r_s \le 1.10$),
which together with the use of recently developed finite-size
correction schemes for QMC energies, allows a significant
reduction of size effects; see appendix \ref{app:sizeef} for
details and discussion.

We performed a 36 CEIMC simulations (not including those
related to the coupling constant integration), the results are
reported in Table III of appendix \ref{app:eos1}. Our
simulations were first equilibrated using effective pair
potentials built from reflected Yukawa functions, which were
chosen such they reproduced the radial distribution functions
of the QMC systems. We then performed 2000-3000 equilibration
steps with CEIMC. Statistics were gathered in the following
5000-15000 steps, the number depending on temperature and
density.

The protons are treated as classical particles in the results
presented in Table III and in the free energy calculations
discussed below. Quantum effects of the protons could be
important at low temperatures at the densities considered in
this work. In order to assess their effect on the thermodynamic
properties, we performed Path Integral Monte Carlo (PIMC)
calculations for the protons on the potential energy surface
defined by the zero temperature RQMC method. This is an
extension of CEIMC to path integral calculations
\cite{Pierleoni06}. The resulting corrections to the energy and
pressure are given in Table I, at T= 2000K for 3 densities. At
this temperature, which is the lowest temperature studied in
this work, the corrections to the pressure are below 1$\%$.

\begin{table}[tbp]
\centering
\begin{tabular}{|c|c|c|}
\hline  $r_s$  &  $\Delta E$ (mHa) &  $\Delta P$ (GPa)   \\
\hline
1.05  &  4.0 (7)  &  7 (3) \\
\hline
1.10  &  3.8 (3)   &  9 (1)  \\
\hline
1.25  &  2.8 (5)  &  5 (1) \\
\hline
\end{tabular}
\label{tab:PIcorr} \caption{Corrections to the energy and
pressure of hydrogen from quantum effects of the protons, from
PIMC simulations with CEIMC, at a temperature of 2000 K:
$\Delta E=(E - E_{classical})/N$ and $\Delta P=P -
P_{classical}$. Errors in parentheses.}
\end{table}

\subsection{Free Energy Integration}

\begin{figure}[tbp]
 \includegraphics[scale=0.65]{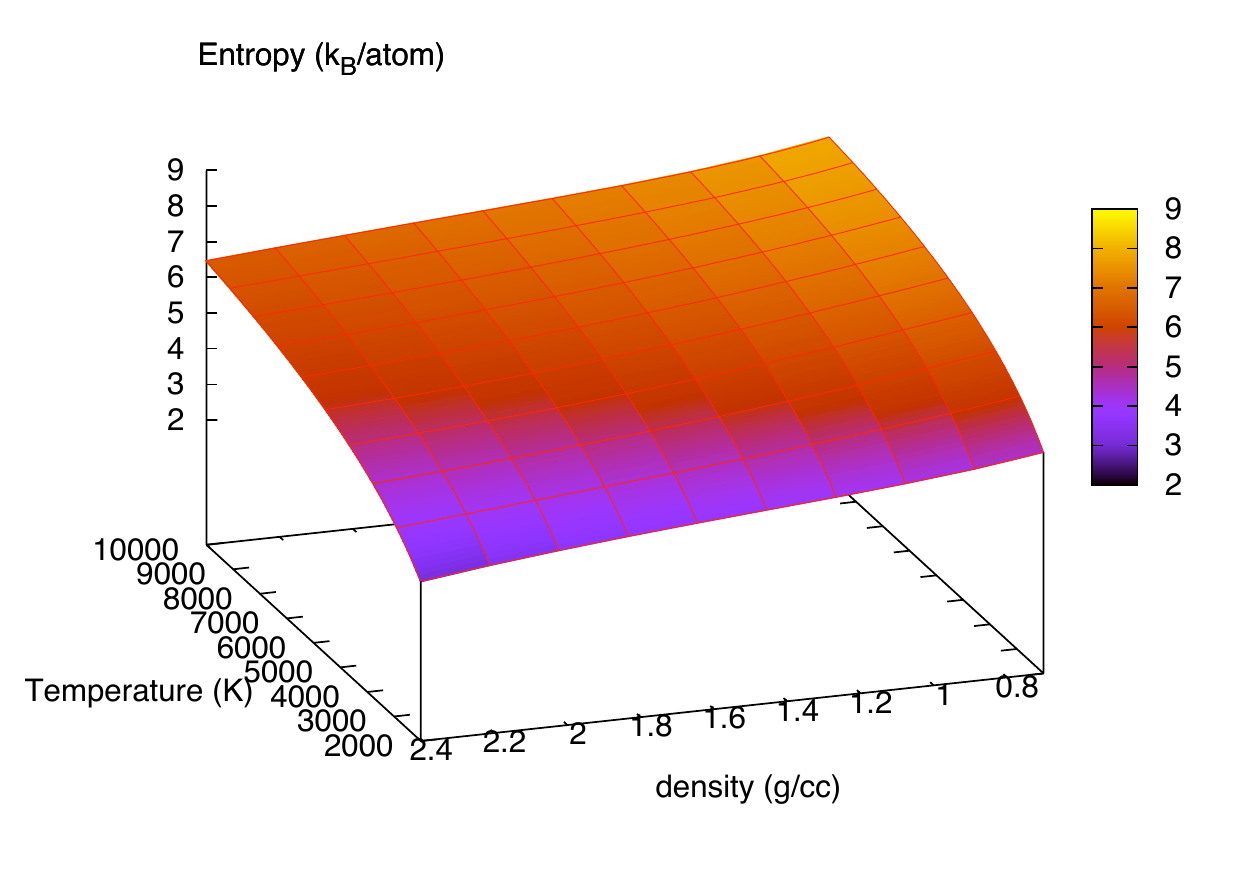}
\caption{\label{fig2} The entropy per atom as a function of
temperature and density. }
\end{figure}

We used Coupling Constant Integration (CCI) \cite{Kirkwood35}
to calculate the free energy of hydrogen at a reference point
chosen as T=6000K and $r_s=1.25$. In CCI a $\lambda$-dependent
potential energy, $V(\lambda)$, is a linear combination of two
different potentials: $V(\lambda)= V_0+ \lambda (V_1-V_0)$. The
difference in free energy between systems $0$ and $1$ is then:
 \begin{equation} F_1 - F_0 = \int_0^1 d\lambda \left< (V_1-V_0) \right>_{V(\lambda)} . \end{equation}
We chose the reference potential to be a reflected Yukawa pair
potential:
 \begin{equation}
V_0(r) =
\begin{cases}
 \frac{e^{-br}}{r} + \frac{e^{-b(L-r)}}{(L-r)} - 4 \frac{e^{-bL/2}}{L} &  r  \le L/2, \\
0 &  r > L/2,
\end{cases}
\end{equation}
where b=2.5 a.u. and L is the length of the simulation cell.
The reflection makes the function and its first derivative
continuous at the cut-off, $r=L/2$. We first used CCI to
calculate the free energy difference between the Yukawa
potential and a system of non-interacting particles. We
performed a second CCI with CEIMC to calculate the free energy
difference between the Yukawa model and the QMC system.
We obtained a value of -0.5737(1) Ha/atom for the
free energy  and 1.98(1)$\ast 10^{-5}$ (Ha/K) for the entropy
at the reference point.

\subsection{Fits to EOS}
The free energy as a function of temperature and density was
obtained by using the functional form:
\begin{equation}
F(T,\rho) = \sum_{i=1}^{4} \sum_{k=1}^{5} c_{ik} \ g_i(T) \ g_k(\rho),
\end{equation}
where $g_i(x) = \{1, x, x^2, x \ln{x}, x^2 \ln{x} \}$. 
We do a least squares fit between the analytical function and the CEIMC data using derivatives of this
expansion:
\begin{eqnarray}
P(T,\rho) & = & \frac{\rho^2}{m_h} \left ( \frac{\partial{F}}{\partial{\rho}} \right )_{T} \\
E(T,\rho) & = & -T^2 \left [ \frac{\partial}{\partial{T}} \left (\frac{F}{T} \right) \right ]_{\rho}  \end{eqnarray}

The free energy fit reproduces the energies and pressures
obtained from the CEIMC simulations to within 0.5$\%$ and
1.5$\%$ respectively, although the average error is much
smaller than this. Figures \ref{fig1} and \ref{fig2} show plots
of the free energy and entropy in the region of the phase
diagram studied. The coefficients of the expansion are given in
Table II.
\begin{table*}[htbp]
\centering
\begin{tabular}{|c|c|c|c|c|}
\hline  k  &  $c_{1k}$  &  $c_{2k}$  & $c_{3k}$  &  $c_{4k}$    \\
\hline
  1   &   -0.529586 $$          &  -2.085591$\cdot10^{-4}$            &  -3.365628$\cdot10^{-9}$      &   2.294411$\cdot10^{-5}$    \\
  2   &   2.227221$\cdot10^{-6}$    &  -1.452601$\cdot10^{-4}$  &  -2.488880$\cdot10^{-9}$   &  1.894880$\cdot10^{-5}$   \\
  3   &   -6.266619$\cdot10^{-5}$  &  4.210279$\cdot10^{-4}$    &  6.174066$\cdot10^{-9}$    &  -5.144879$\cdot10^{-5}$  \\
  4   &  9.977346$\cdot10^{-2}$  &  -6.220508$\cdot10^{-4}$    & -8.564851$\cdot10^{-9}$    &  7.499558$\cdot10^{-5}$  \\
  5   &   -1.437627$\cdot10^{-2}$  & -9.867541$\cdot10^{-5}$    &  -1.598083$\cdot10^{-9}$   & 1.225739$\cdot10^{-5}$  \\
\hline
\end{tabular}
\label{tab:FEcoeffs} \caption{Coefficients of the expansion of
the free energy; energy in Hartree/atom, temperature in K and
density in g $cm^{-3}$}
\end{table*}

 \begin{figure}[h]
 \includegraphics[scale=0.8]{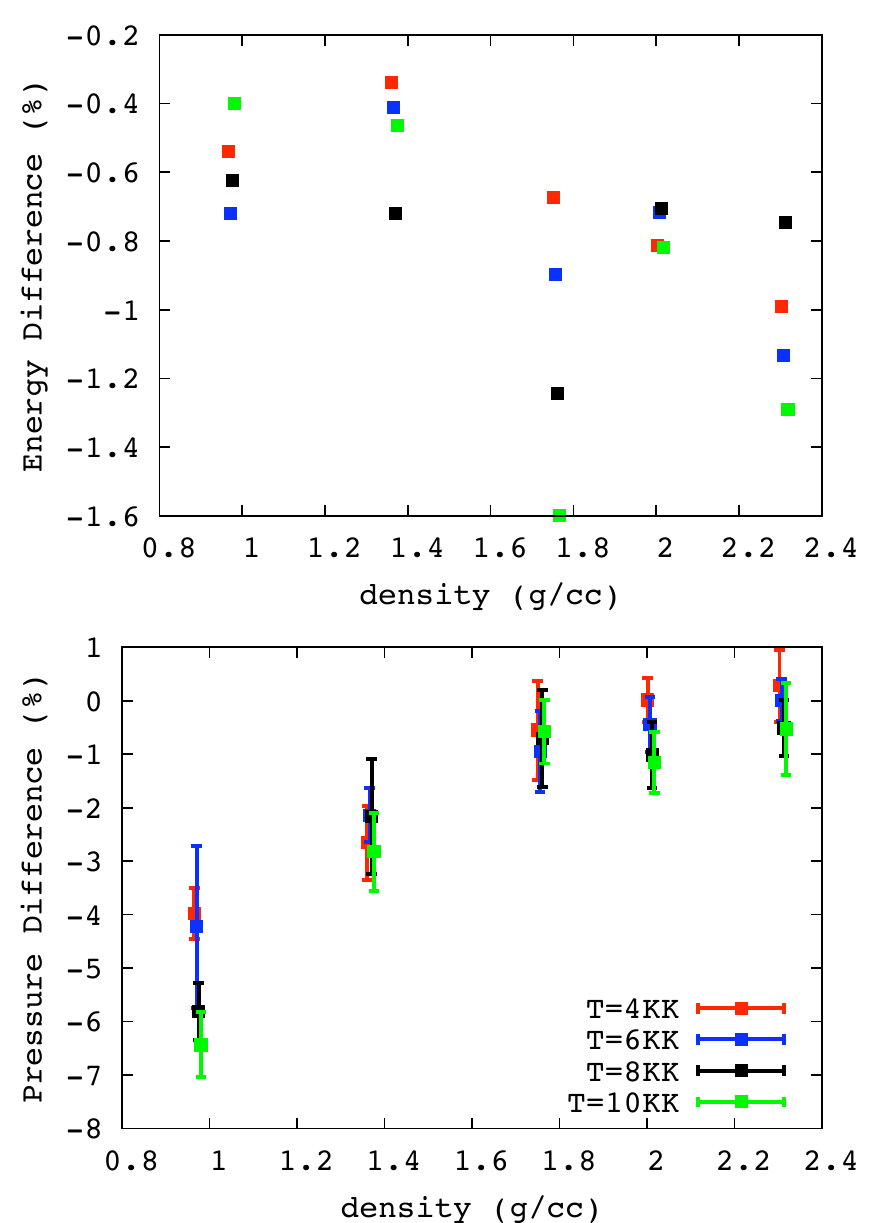}
 \caption{\label{fig3} Comparison of EOS data with DFT-based BOMD simulations. }
\end{figure}

\section{Comparison with other methods}
\label{sec:Compare}

Figure \ref{fig3} shows a comparison of the pressure and energy
between CEIMC simulations and DFT based Born-Oppenheimer
Molecular Dynamics (BOMD) simulations.
The BOMD simulations were performed in the NVT-ensemble (with a
weakly coupled Berendsen thermostat) using the Qbox code
\footnote{http://eslab.ucdavis.edu/software/qbox}. We used the
Perdew-Burke-Ernzerhof (PBE) exchange-correlation functional
and a Hamman type \cite{Hamman89} local pseudopotential with a
core radius of $r_c$ = 0.3 $a.u.$ to represent hydrogen. The
simulations were performed with 250 hydrogen atoms using a
plane-wave cutoff of 90 Ry (115 Ry for rs $\ge$ 1.10)with
periodic boundary conditions ($\Gamma$ point). Corrections to
the EOS were added to extrapolate results to infinite cutoff
and to account for the Brillouin zone integration. To do this
we studied 15-20 statistically independent static
configurations at each density by using a 4x4x4 grid of
k-points with a plane-wave cutoff of at least 300 Ry. See ref.
\cite{Morales09} for additional details of the DFT simulations.

\begin{figure}
 \includegraphics[scale=0.45]{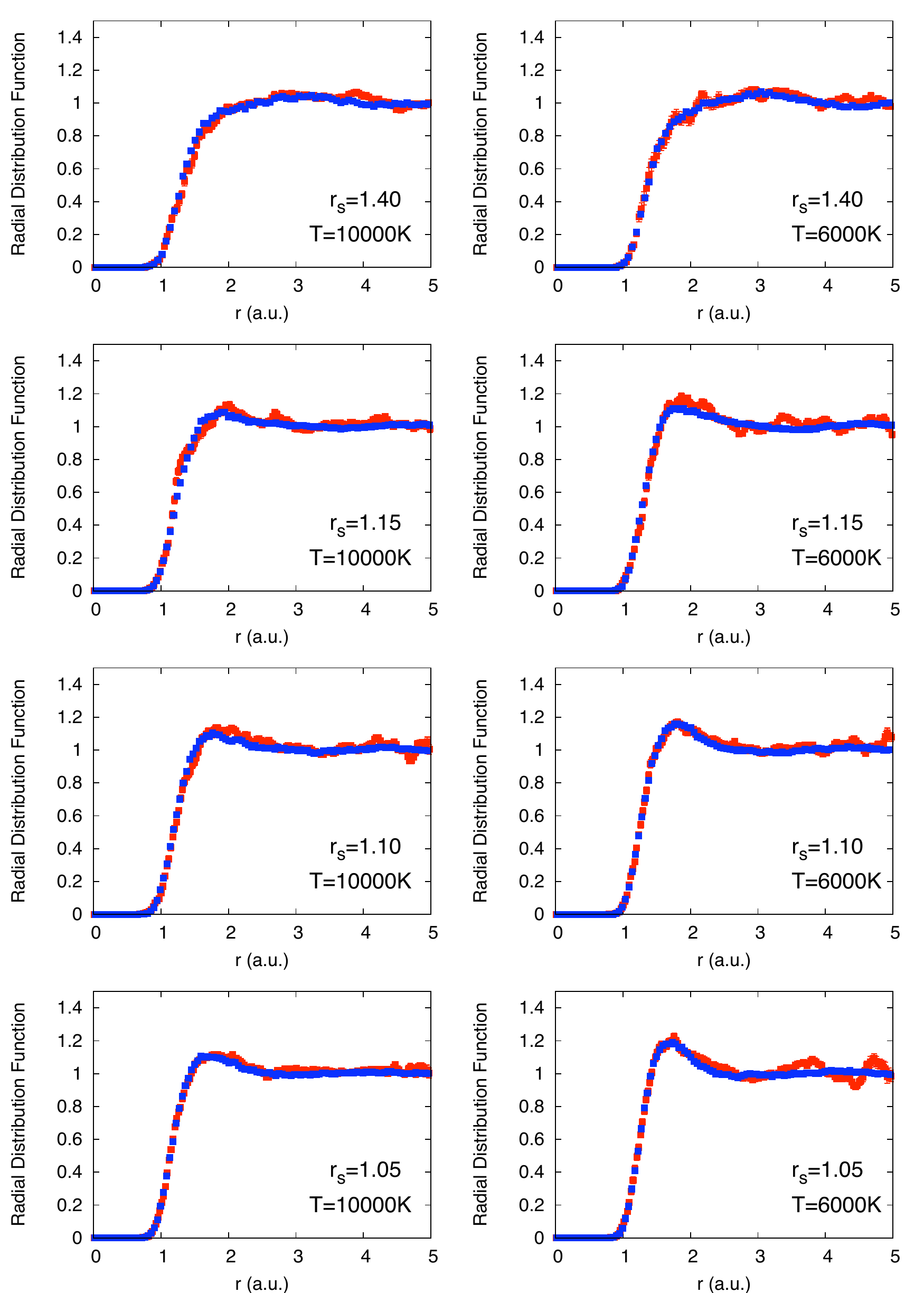}
 \caption{\label{fig4} Comparison of radial distribution functions between BOMD (blue)  and CEIMC (red).}
\end{figure}

\begin{figure}
 \centering
 \includegraphics[scale=0.9]{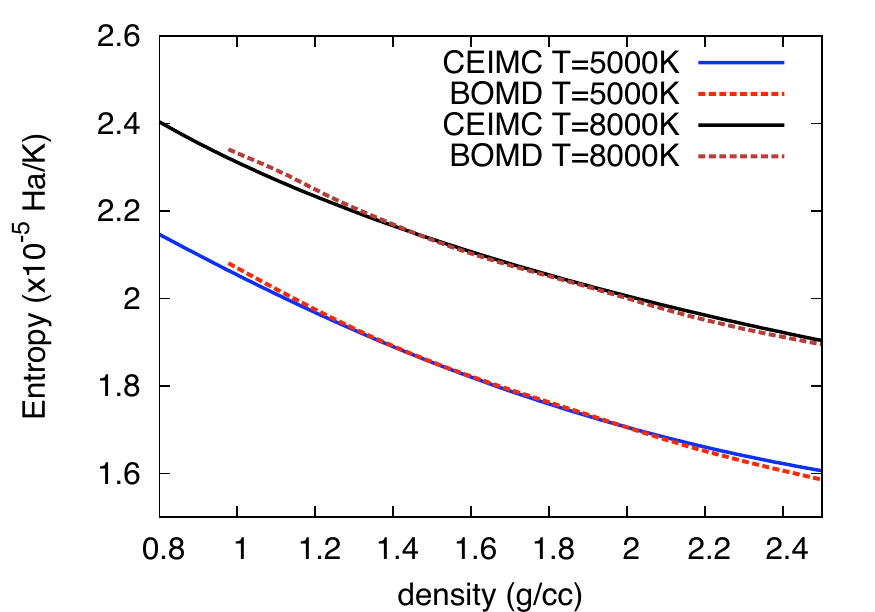}
 \caption{\label{fig5} The entropy/atom at 5000K and 8000K as determined with BOMD and CEIMC. }
\end{figure}

There is a good agreement between the two methods, especially
at higher densities where the difference in pressure is within
error bars. At lower densities, the pressure difference
increases reaching an average of approximately 5$\%$ close to
the dissociation regime. There is less reason to expect exact
agreement of the energies calculated since the DFT calculations
use pseudopotentials as well as approximate
exchange-correlation functionals which can modify the zero of
energy. However, the temperature and density dependence is well
reproduced with an almost uniform energy difference of 0.8$\%$
on the entire phase diagram. Figure \ref{fig4} shows a
comparison of the proton-proton distribution function for
several thermodynamic conditions. The agreement between the two
methods is again remarkable. The structure of the liquid is
reproduced by the DFT simulations quite accurately, even the
short range correlation peak that develops at the lower
temperatures and higher densities. Figure \ref{fig5} shows a
comparison of the entropy as a function of density along
several isotherms. We can see that for densities beyond
$\rho=1.4 g/cm^3$, the entropies obtained with the two methods
are undistinguishable. In general, we obtain very good agreement
between the two methods for pressures beyond 600 GPa. At lower
pressures, the agreement is not perfect, but still very good,
with BOMD predicting a slightly higher entropy than CEIMC. We
are currently expanding our calculations to lower density to
provide an additional benchmark of the DFT method close to the
molecular dissociation region.

\begin{figure}
 \centering
 \includegraphics[scale=0.90]{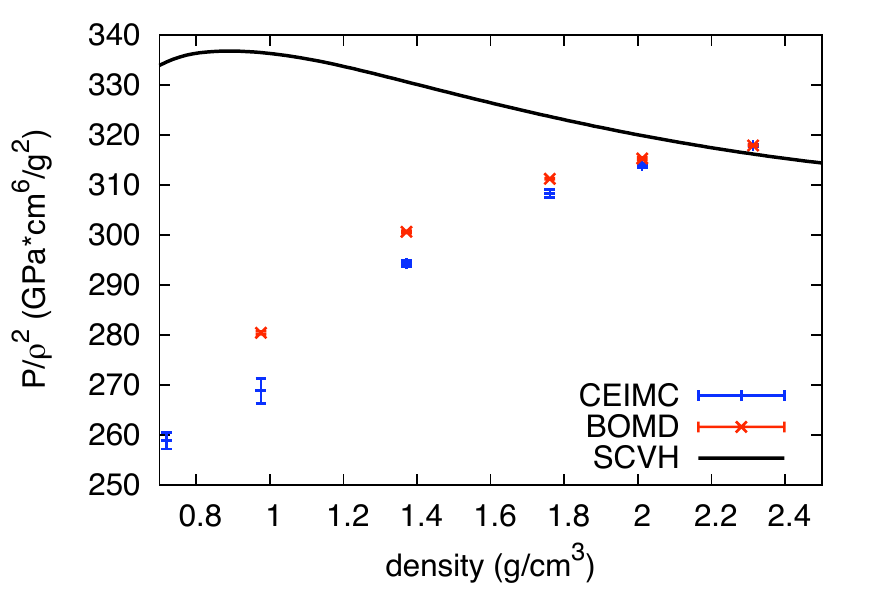}
 \caption{\label{fig6} Comparison of pressure (divided by $\rho^2$) as a function of density computed with SCVH, BOMD and CEIMC at T=6000 K. }
\end{figure}

Figure \ref{fig6} shows a comparison of the pressure as a
function of density obtained with CEIMC and BOMC and the SCVH
equation of state at T=6000 K.
At pressures well below the dissociation regime, SCVH produces
very good results, but at higher densities 
the model can not capture all the features which result from
strong atomic coupling; it predicts a qualitatively different
behavior with higher pressures at lower densities.

\section{Conclusions}
\label{sec:Conclusion}

In summary, we performed a comprehensive study of the free
energy and the equation of state of warm dense liquid hydrogen
in its atomic phase using energies computed with quantum Monte
Carlo methods. We provide a fit to the free energy which can be
used as input to models of Jovian planets or in the formulation
of more accurate chemical models. Given the current status of
DFT and its possible limitations at these conditions, it is
crucial to benchmark its predictions against more accurate
methods. We provide such a critical test. Our results indicate
that DFT-based BOMD simulations provide a very good description
of both thermodynamic and structural properties of hydrogen for
the studied conditions. The equation of state of SCVH, used in
the study of planetary interiors for more than a decade, is
shown to produce inaccurate results in the atomic regime. This
suggests that planetary models should be reinvestigated with a
more accurate equations of state, such as the one presented in
this work.

\appendix
\section{Equation of State Table}
\label{app:eos1}

Table III gives the energy per atom and pressure as calculated
by CEIMC on a grid in temperature and density. Finite size
corrections (see next section) are included in the reported
results, as well as corrections to the pressure from non-zero
time step and finite projection time in RQMC. The latter
corrections are negligible in the case of total energies.

\begin{table}[h]
\centering
\begin{tabular}{cccc}
\hline
\hline
T (K)  & r$_s$  &  Energy (Ha) & Pressure(GPa)   \\
\hline
2000   &   1.05   &  -0.3846 (3)    &  1576 (2)   \\
3000   &   1.05   &  -0.3777 (2)   &  1607 (1)  \\
4000   &   1.05   &   -0.3707 (5)   &   1640 (3)  \\
6000   &   1.05   &  -0.3569 (4)  &   1701 (2)   \\
8000   &   1.05  &  -0.3458 (4)  &   1753 (2)  \\
10000   &   1.05   &   -0.3316 (8)  &   1814 (4)  \\
\hline
2000   &   1.10   &  -0.4170 (2)    &  1157 (2)   \\
3000   &   1.10   &  -0.4097 (2)   &  1190 (1)  \\
4000   &   1.10   &   -0.4026 (3)   &   1219 (1)  \\
6000   &   1.10   &  -0.3898 (4)  &   1270 (2)   \\
8000   &   1.10  &  -0.3777 (5)  &   1315 (2)  \\
10000   &   1.10   &   -0.3660 (4)  &   1362 (2)  \\
\hline
2000   &   1.15   &  -0.4419 (2)    &  861 (1)   \\
3000   &   1.15   &  -0.4356 (3)   &  883 (2)  \\
4000   &   1.15   &   -0.4285 (7)   &   911 (3)  \\
6000   &   1.15   &  -0.4151 (6)  &   956 (3)   \\
8000   &   1.15   &  -0.4018 (9)  &   1003 (3)  \\
10000   &   1.15   &   -0.3891 (8)  &   1048 (2)  \\
\hline
2000   &   1.25   &   -0.4790 (2)    &  485 (1)   \\
3000   &   1.25   &    -0.4721 (2)   &  504 (1)  \\
4000   &   1.25   &   -0.4673 (4)   &  517 (2)  \\
6000   &   1.25   &  -0.4549 (6)  &   556 (1)   \\
8000   &   1.25   &  -0.4419 (7)  &   590 (3)  \\
10000   &   1.25   &  -0.4324 (7)  &   619 (2)  \\
\hline
2000   &   1.40   &   -0.5117 (4)  &   214 (1)  \\
3000   &   1.40   &   -0.5057 (2)  &   222 (1)  \\
4000   &   1.40   &   -0.4993 (5)  &   234 (1)  \\
6000   &   1.40   &   -0.4869 (5)  &   257 (3)  \\
8000   &   1.40   &   -0.4767 (5)  &   277 (1)  \\
10000   &   1.40   &   -0.4674 (4)  &   297 (1)  \\
\hline
2000    &   1.55   &  -0.5330 (2)  & 111 (1)  \\
3000    &   1.55   &  -0.5230 (2)  &  105 (1)  \\
4000    &   1.55   &  -0.5157 (3)  &  117 (1) \\
6000    &   1.55   &  -0.5027 (4)  &  134 (1) \\
8000    &   1.55   &   -0.4938 (2)  & 143 (1)  \\
10000    &   1.55   &  -0.4831 (6)  & 163 (2)  \\
\hline
\end{tabular}
\label{table:eostable1} \caption{Energy/atom and pressure as
calculated with CEIMC. Statistical errors are given in
parentheses.}
\end{table}

\section{Finite Size Effects}
\label{app:sizeef}

Due to the high computational demands of QMC, our simulations
are restricted to systems with  128 atoms or fewer. Many
techniques have been developed in order to obtain useful
results with finite systems. In this work we use TABC (the
generalization of Brillouin zone integration to many-body
quantum systems) to eliminate shell effects in the kinetic
energy of metallic systems.  Twisted boundary conditions when
an electron wraps around the simulation box are defined by:
\begin{equation}
\Psi_{\theta}(...,\vec{r_j}+\vec{L},...) = e^{i \theta} \Psi_{\theta}(...,\vec{r_j},...).
\end{equation}
Observables are then averaged over the all twist vectors,
similar to one-body theories:
\begin{equation}
< \hat A > = \int_{-\pi}^{\pi} \frac{d \vec{\theta}}{(2 \pi)^3}
< \Psi_{\theta} | \hat A | \Psi_{\theta}  >.
\end{equation}
This has been shown to restore the $1/N$ dependence of the
energy in QMC calculations, absent when PBC are used.

As first shown by Chiesa et.al. \cite{Chiesa06}, most of the
remaining finite size errors in the potential and kinetic
energies of QMC simulations come from discretization errors
induced by the use of PBC. To see this, notice that we can
write the potential energy of a system of N electrons as:
\begin{equation}
<\hat{V}> = \frac{1}{2 \Omega} \sum_{\vec{k} \neq 0} v(\vec{k}) [S_{N}(\vec{k}) - 2N] ,
\end{equation}
where $\Omega$ is the volume of the supercell, $v(\vec{k})$ is
the Fourier transform of the Coulomb potential,
$S_{N}(\vec{k})$ =  $< \rho(\vec{k}) \rho(-\vec{k}) >$ is the 
structure factor, $\rho(\vec{k}) = \sum_i z_i e^{i
\vec{k} \cdot \vec{r_i}}$ \footnote{The sum over {\it i} includes both protons and electrons.} and $\vec{k}$ is the set of lattice
vectors in reciprocal space of the supercell. As we approach
the thermodynamic limit $N\rightarrow \infty$, the structure
factor converges ($S_N(\vec{k}) \rightarrow
S_{\infty}(\vec{k})$) and the sum becomes an integral:
$\frac{1}{V} \sum_{\vec{k} \neq 0} \rightarrow \int \frac{d
\vec{k}}{(2 \pi)^3}$. If we assume that the structure factor is
essentially converged for some finite number of atoms, then
most of the finite size error in the potential energy comes from the
omission of the $\vec{k}=0$ term in the sum. This can be
estimated using the Poisson summation formula:
\begin{equation}
\int \frac{d \vec{k}}{(2 \pi)^3} \hat{\eta}(\vec{k})  - \sum_{\vec{k} \neq 0} \hat{\eta}(\vec{k}) = \eta(0) - \sum_{\vec{L}} \eta(\vec{L}).
\end{equation}
We know from the RPA, exact in the limit of $\vec{k}
\rightarrow 0$, that $S(\vec{k}) \approx k^2$ as $k \rightarrow
0$. The leading order correction to the potential energy is:
\begin{equation}
\delta V = \frac{3}{2 N r_s^3} \lim_{\vec{k} \rightarrow 0} \frac{S(\vec{k})}{k^2}.
\end{equation}
By estimating the structure factor for small k during our
simulation and extrapolating its behavior to k=0, we obtain the
desired corrections. In the case of the kinetic energy,
following a similar argument we obtain for the correction to
second order \cite{Chiesa06,Drummond08}:
\begin{eqnarray}
\delta K & = & \int \frac{d \vec{k}}{(2 \pi)^3} u(\vec{k}) k^2 - \sum_{\vec{k} \neq 0} u(\vec{k}) k^2 \nonumber \\
 & = & \frac{\sqrt{3}}{4 N r_s^{3/2}} - \frac{5.264}{2\pi r_s^2 (2N)^{4/3}},
\end{eqnarray}
where $u(\vec{k})$ is the electron-electron Jastrow; we used
its RPA form.

\begin{figure}
 \includegraphics[scale=0.8]{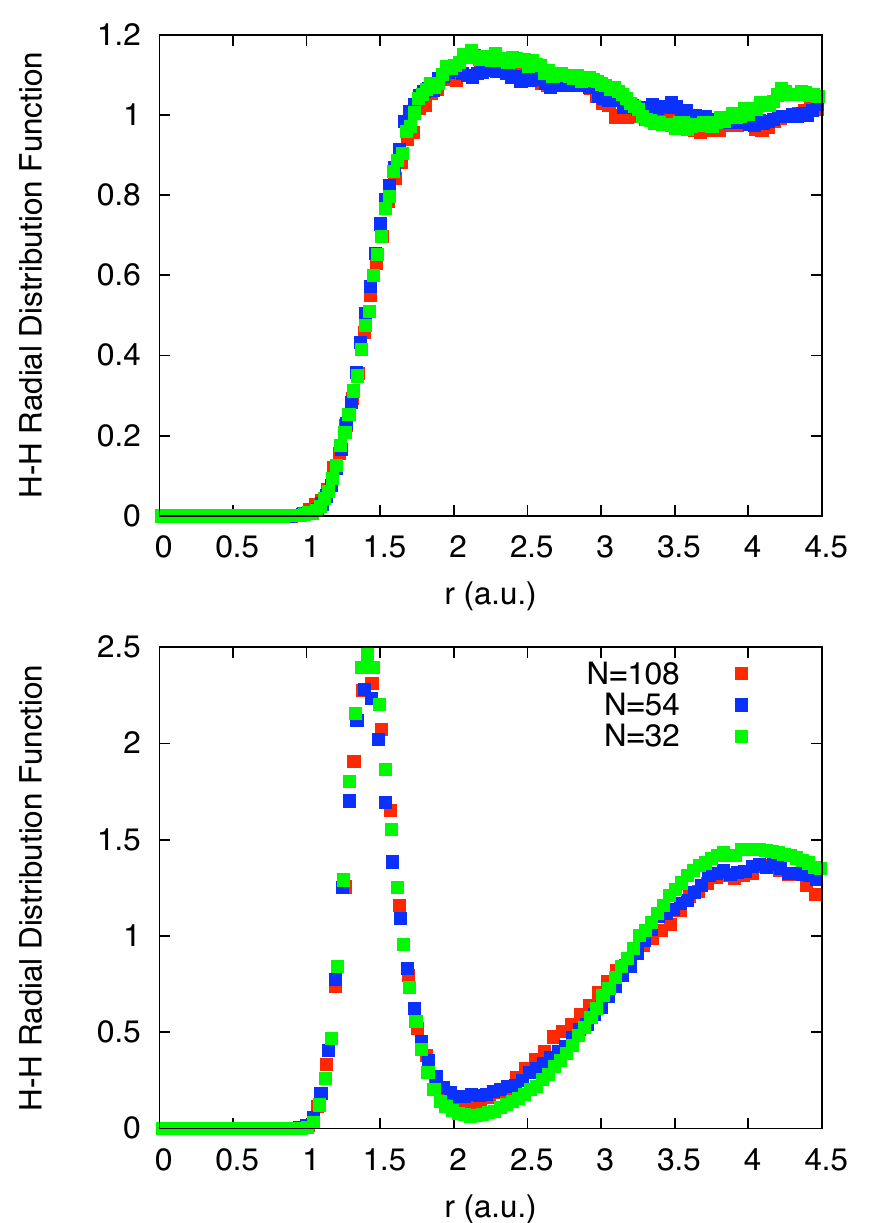}
\caption{\label{fig7} Comparison of proton-proton radial
distribution functions between systems with different number of
electrons. In the upper panel, the temperature is 4000K and
r$_s$=1.25. In the lower panel, the temperature is 3000K and
r$_s$=1.85. Electronic energies in CEIMC were calculated using
Variational Monte Carlo.}
\end{figure}

To check the finite-size corrections for dense hydrogen, we
performed simulations with 32, 54 and 108 atoms at $r_s$=1.85
and $r_s$=1.25. We used TABC  with Variational Monte Carlo
energies with 108 twists for the 32 atom system and 32 twists
for the 54 and 108 atoms.  Figure \ref{fig7} shows a comparison
of the radial distribution functions between the systems with
different number of atoms, for the two densities studied. As
can be seen, the agreement between the 3 simulations is very
good, with no noticeable difference between the systems with 54
and 108 atoms. Using the fact that TABC restores the 1/N
dependence of the properties, we can compare the results of the
size correction formulas with a 1/N extrapolation. Table IV
shows a comparison of finite size corrections taking the system
with 54 atoms as the reference. As can be seen, the correction
for the lower density system agrees very well with the
extrapolated value. In  the case of the higher density system,
the agreement is less good but still acceptable. We attribute the disagreement 
at higher densities to the differences in TABC used for the systems with
different sizes.

\begin{table}[h]
\centering
\begin{tabular}{|c||c|c||c|c||}
\hline
 $r_s$  & \multicolumn{2}{|c||}{Energy (mHa)} &   \multicolumn{2}{|c||}{Pressure (GPa)}  \\
 \hline
              &  $\Delta E_N$  & $\Delta E_S$ & $\Delta P_N$  & $\Delta P_S$  \\
 \hline
     1.25    &  7.7   &  10.0  &  10.8  &  17.6  \\
     \hline
     1.85   &   5.4   &    5.5   &   3.3   &   3.1  \\
     \hline
 \end{tabular}
\label{tab:fscorr}
\caption{Comparison of finite-size corrections between a size extrapolation ($\Delta E_N$,$\Delta P_N$) and formulas  B5 and B6 ($ \Delta E_S$,$ \Delta P_S$).}
\end{table}

\section{Details of RQMC calculations}
\label{app:convergence}

As mentioned in the text, the parameters of the RQMC
calculations were varied with density to obtain uniform
accuracy over the whole phase diagram. Table V shows the time
steps and projection times used in the simulations for the
different densities studied. As is well known, total energies
converge faster as a function of imaginary time because their
convergence error is second order with respect to the trial
wavefunction. The kinetic and potential energies, and hence the
pressure, are only first order, which means that longer
projection times are needed to obtain converged results. During
the course of the simulations, we only require accurate
energies, so a smaller projection time is used; one that is
insufficient for accurate pressures. In order to obtain
converged results for the pressure, we calculated a correction
for the finite projection time and non-zero time step  by
studying approximately a dozen protonic configurations at each
density. The corrections were found to be independent of the
precise proton configuration but density dependent. The
corrections to the total energy are negligible within errors.
\\
\\
\begin{table}[h]
\centering
\begin{tabular}{|c|c|c|}
\hline
 $r_s$  & projection time $(a.u.)^{-1}$  &  time step (a.u.) \\
 \hline
     1.05    &  0.456  &   0.008   \\
     \hline
     1.10   &  0.504   &   0.008   \\
     \hline
      1.15  &  0.550   &   0.01   \\
     \hline
      1.25  &  0.660   &   0.012   \\
     \hline
      1.40  &  0.732   &   0.012   \\
     \hline
      1.55  &  0.975  &   0.015   \\
     \hline
 \end{tabular}
\label{tab:rqmcdetails}
\caption{Projection time and time step used in the RQMC calculations.}
\end{table}

\begin{acknowledgments}
We thank M. Holzmann for useful discussions on the finite size
corrections. MAM acknowledges support from the SSGF program,
DOE grant  DE-FG52-06NA26170 and computer time from the DOE
INCITE program. C.P. thanks the Institute of Condensed Matter Theory at the University of Illinois at Urbana-Champaign for a short term visit, and acknowledges financial support from Ministero dell'Universitˆ e della Ricerca, Grant PRIN2007.
\end{acknowledgments}

\end{document}